\documentclass{iau}
\usepackage{graphicx}
\def\beq{\begin{equation}}
\def\enq{\end{equation}}

\title[IAUS291.~~Superslow pulsation X-ray pulsars] 
{The superslow pulsation X-ray pulsars in high mass X-ray binaries} 

\author[W.~Wang]  
{Wei Wang
 }

\affiliation{National Astronomical Observatories, Chinese Academy of Sciences, Beijing 100012, China \\ email: {wangwei@bao.ac.cn}
}

\pubyear{2012}
\volume{291}  
\jname{\mbox{Neutron Stars and Pulsars: Challenges and Opportunities after 80 years}}
\editors{J. van Leeuwen, ed.}
\begin{document}

\maketitle

\begin{abstract}
There exists a special class of X-ray pulsars that exhibit very slow
pulsation of $P_{\rm spin}>1000$\,s in the high mass X-ray binaries
(HMXBs). We have studied the temporal and spectral properties of these
superslow pulsation neutron star binaries in hard X-ray bands with
\mbox{INTEGRAL} observations. Long-term monitoring observations find spin
period evolution of two sources: spin-down trend for 4U 2206+54
($P_{\rm spin}\sim 5560$\,s with $\dot{P}_{\rm spin}\sim 4.9\times
10^{-7}$\,s s$^{-1}$) and long-term spin-up trend for 2S 0114+65
($P_{\rm spin}\sim 9600$\,s with $\dot{P}_{\rm spin}\sim -1\times
10^{-6}$\,s s$^{-1}$) in the last 20 years. A Be X-ray transient, SXP~1062 ($P_{\rm spin}\sim 1062$\,s), also showed a fast spin-down rate of
$\dot{P}_{\rm spin}\sim 3\times 10^{-6}$\,s s$^{-1}$ during an
outburst. These superslow pulsation neutron stars cannot be produced
in the standard X-ray binary evolution model unless the neutron
star has a much stronger surface magnetic field ($B>10^{14}$\,G). The
physical origin of the superslow spin period is still unclear. The
possible origin and evolution channels of the superslow pulsation X-ray
pulsars are discussed. Superslow pulsation X-ray pulsars could be
younger X-ray binary systems, still in the fast evolution
phase preceding the final equilibrium state. Alternatively, they could be a
new class of neutron star system $-$ accreting magnetars.

\end{abstract}


\firstsection 
\section{Introduction}

Recent X-ray observations discovered some superslow pulsation neutron
star binaries with $P_{\rm spin}>1000$ s. In Fig.\ 1, the Corbet
diagram for high mass X-ray binaries shows four superslow pulsation
X-ray pulsars: 4U 2206+54 with $P_{\rm spin}\sim 5560$\,s (Wang 2009,
2010; Reig et al.\ 2009)  and an orbital period of 19.12 days (Wang
2009); 2S 0114+65 with $P_{\rm spin}\sim 9600$\,s (Wang 2011) and an
orbital period of 11.59 days (Crampton et al.\ 1985); IGR J16418-4532
with $P_{\rm spin}\sim 1246$\,s and $P_{\rm orb}\sim 3.7$ days (Walter
et al.\ 2006); and SXP~1062 with $P_{\rm spin}\sim 1062$\,s and $P_{\rm orb}\sim 300$ days (Haberl et al.\ 2012). In addition, other possible
superslow X-ray pulsar candidates were reported recently: 1E
161348-5055 in the young supernova remnant RCW 103 ($P_{\rm
spin}\sim 6.67$\,hr, De Luca et al.\ 2006), and two wind-accretion
symbiotic low mass X-ray binaries 4U 1954+319 ($P_{\rm spin}\sim 5$\,hr, Mattana et al.\ 2006) and IGR J16358-4724 ($P_{\rm spin}\sim 1.5$\,hr, Patel et al.\ 2004).

\begin{figure}
\begin{minipage}[b]{0.66\textwidth}%
\centering
\includegraphics[angle=0,width=1.1\textwidth]{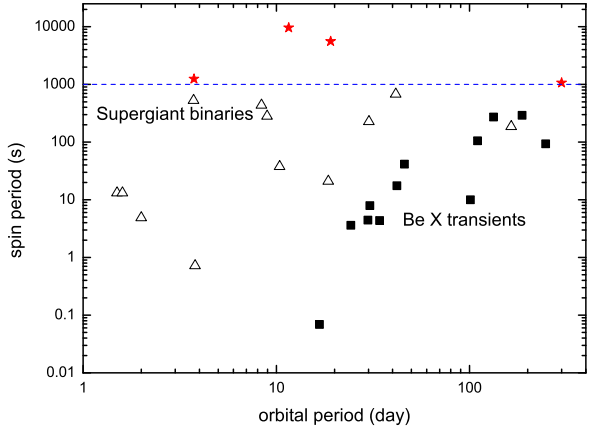}%
\vspace{-10mm}%
\end{minipage}%
\begin{minipage}[b]{0.33\textwidth}%
\caption{The $P_{\rm spin} - P_{\rm orb}$ diagram for high mass X-ray binaries. }
\end{minipage}%
\vspace{4mm}%
\end{figure}

The spin period evolution of the new-born neutron star generally
undergoes three states (Bhattacharya \& van den Heuvel 1991): an {\em
  ejector state} in which neutron star spins down through the
conventional spin-powered pulsar energy-loss mechanisms; a {\em
  propeller state} in which spin period decreases by means of
interaction between the neutron star magnetosphere and stellar wind of
the companion; and an {\em accretor state} in which the spin period of neutron
star reaches a critical value, and the neutron star begins to accrete
materials on to the surface, then switches on as the X-ray pulsar. The
critical period is defined by equating the corotational radius of the
neutron star to the magnetospheric radius, which induces the longest
period of several hundred seconds but less than $\sim$ 1000 s for the
neutron star of magnetic field $B<B_{\rm cr}=4.4\times 10^{13}$
G. Then what channels produce the superlong spin period higher than 1000\,s? Thus, detailed studies of these superslow X-ray pulsars will help us to understand the evolution of neutron star binaries and physical nature of these sources.

\section{Temporal and spectral properties of superslow pulsation X-ray pulsars}

With the long-term INTEGRAL monitoring observations, we derived the orbital phase-resolved spectral properties for two superslow pulsation X-ray pulsars 4U 2206+54 and 2S 0114+65 (Fig.\ 2). The spectra are fitted with the absorbed power-law model plus high energy cut-off. The spectral variations in both two sources show a common property. There exist anti-correlations between the flux and hydrogen column density/photon index, i.e., a lower column density and harder spectrum around maximum of X-ray flux. These spectral behaviour over the orbital phase suggested that they should belong to highly obscured X-ray binary systems.

\begin{figure}[b]
\centering%
\includegraphics[angle=0,width=4.75cm]{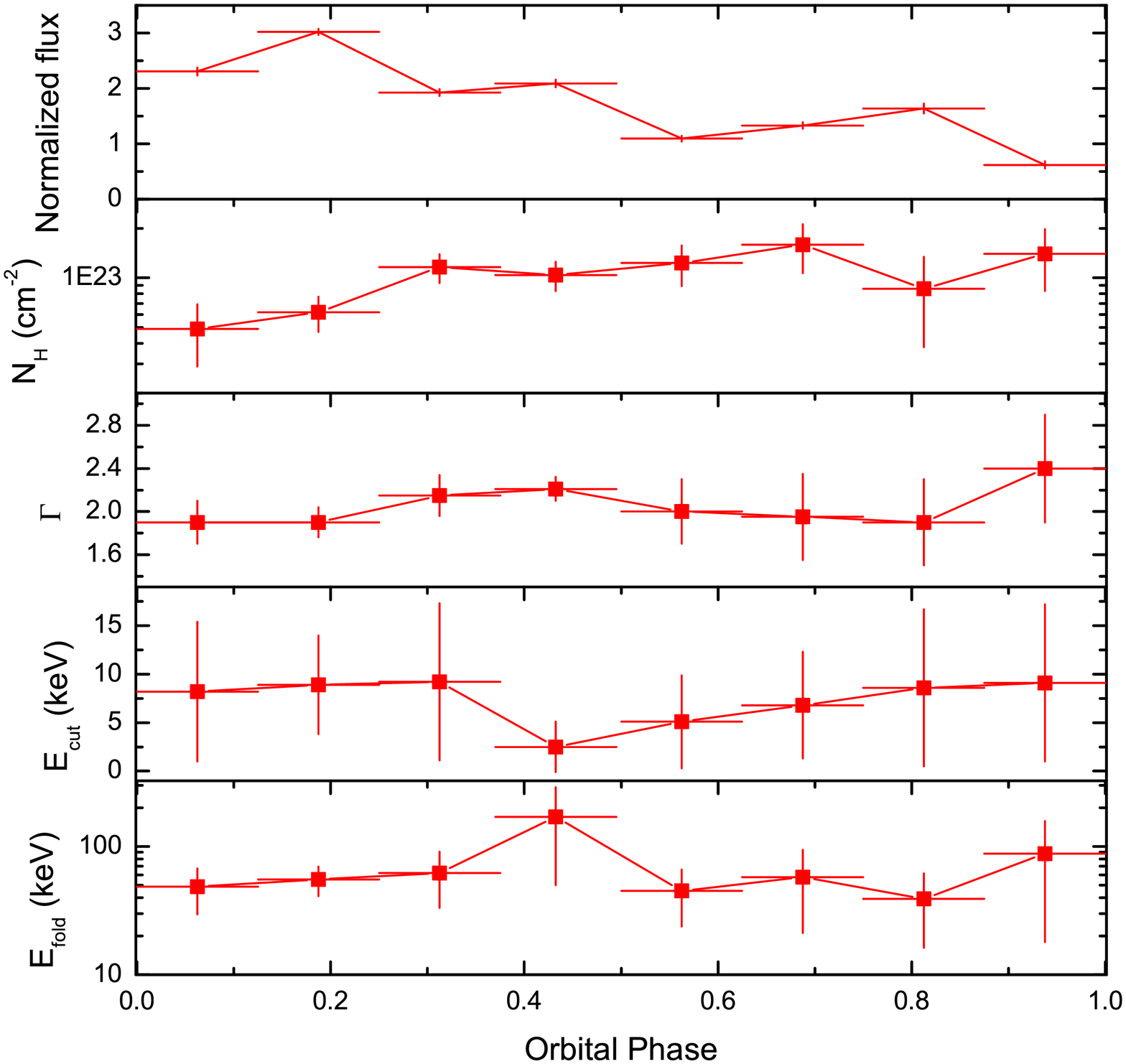}\hspace{-5mm}%
\vspace{-5mm}\includegraphics[angle=0,width=4.75cm]{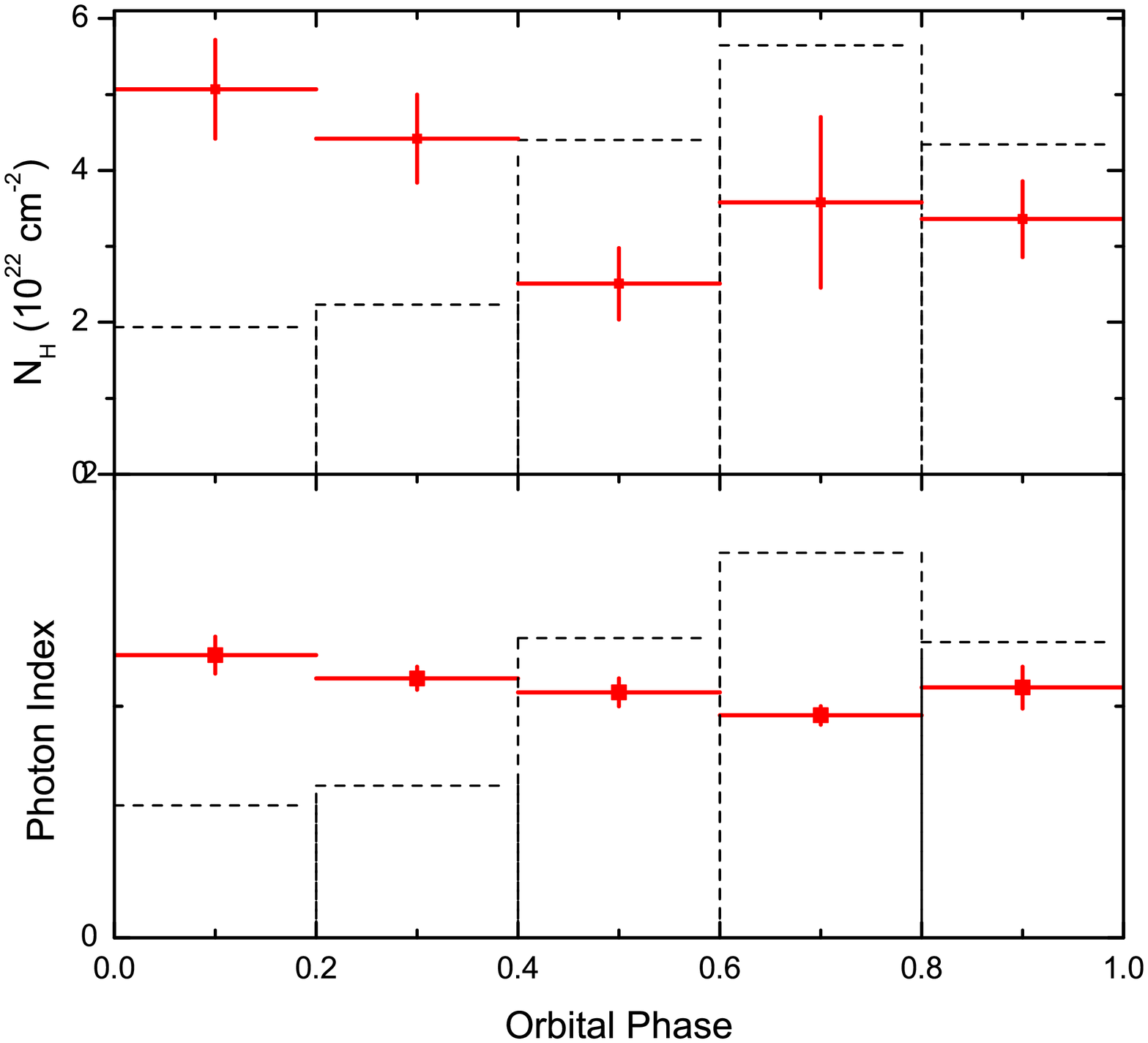}\hspace{-5mm}\vspace{5mm}%
\includegraphics[angle=0,width=4.75cm]{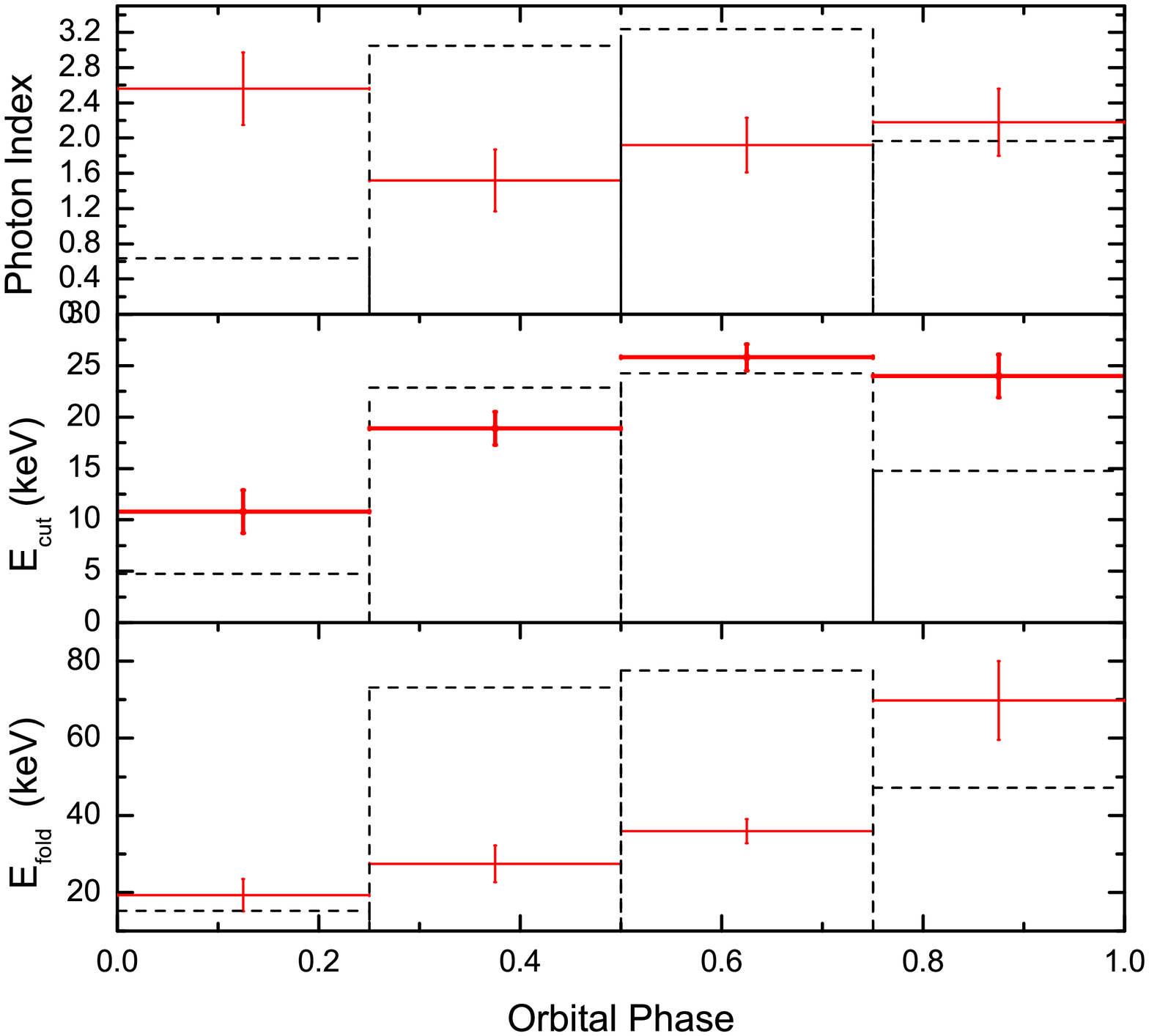}%
\caption{Spectral property variations of 4U 2206+54 (left, Wang 2012) and 2S 0114+65 (middle from RXTE data (Farrell et al.\ 2008) and right from IBIS data, Wang 2011) over orbital phases. }
\end{figure}

In addition, we detected two cyclotron absorption lines at $\sim 30$ keV and $60$ keV in 4U 2206+54 during an active state (Wang 2009), suggesting a magnetized neutron star with the magnetic field of $\sim 3\times 10^{12}$ G located in the binary if assuming the electron absorption case. Unfortunately, we have not found evidence for the magnetic neutron star in 2S 0114+65 with different observations (Wang 2011). While a high energy tail was discovered in the X-ray spectrum of 2S 0114+65 (Wang 2011). With detailed studies show that high column density may
lead to the disappearance of the hard X-ray tails in the spectra: when the derived values of column
density are higher than $\sim 3\times 10^{22}$ cm$^{-2}$, no hard
X-ray tails are detected. How to produce the hard X-ray tails above
70 keV for accreting neutron stars in high mass X-ray binaries
especially in the wind-fed accretion systems is unclear. It is
possible that hot corona exists near neutron stars for wind-fed
accretion systems like 2S 014+65; and the dense accretion
materials or strong winds prevent the formation of hot corona or
depress the comptonization effects.

\begin{figure}
\centering%
\hspace{-5mm}\includegraphics[angle=0, width=7.2cm]{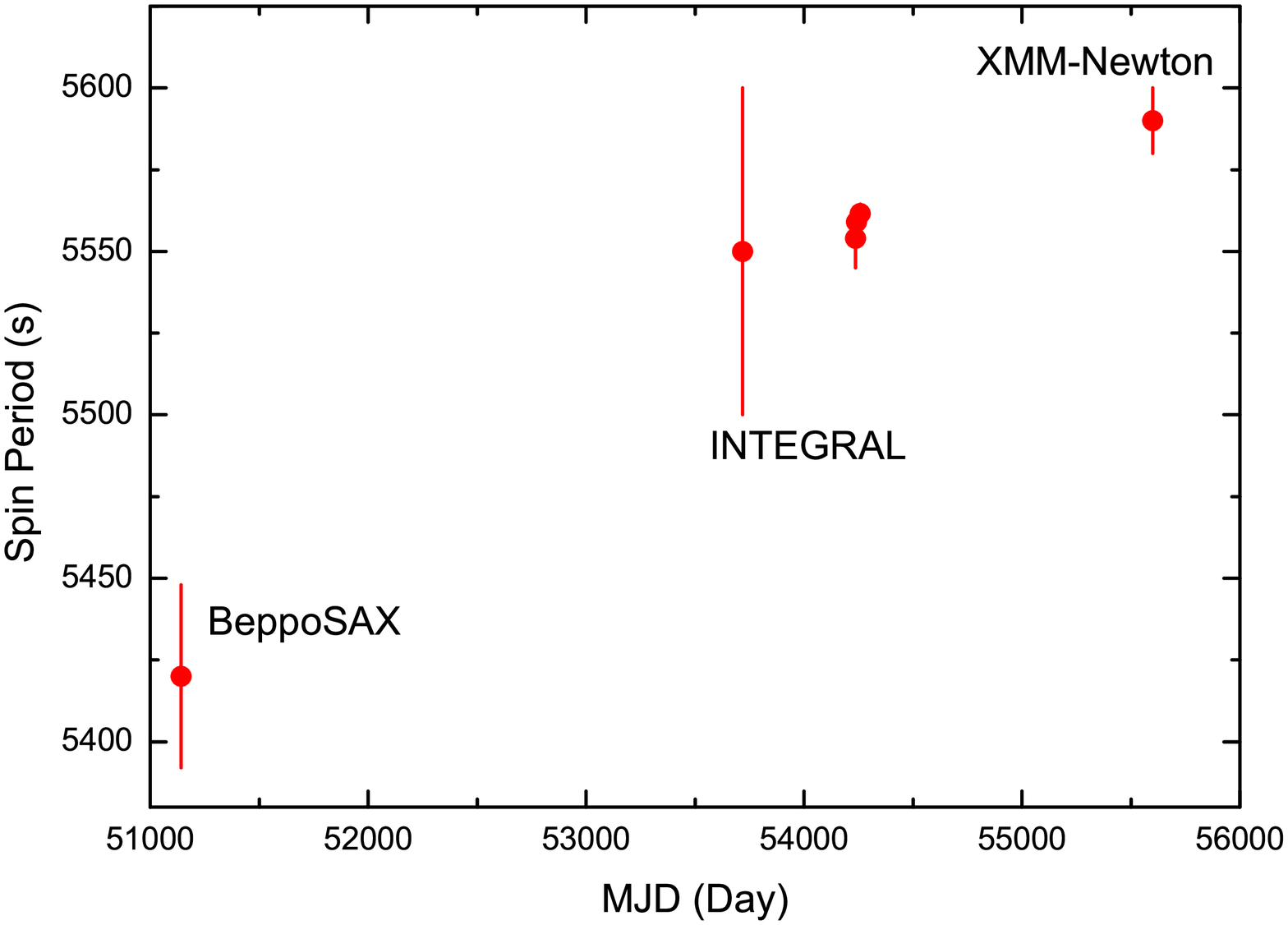}%
\hspace{-10mm}\includegraphics[angle=0,width=7.2cm]{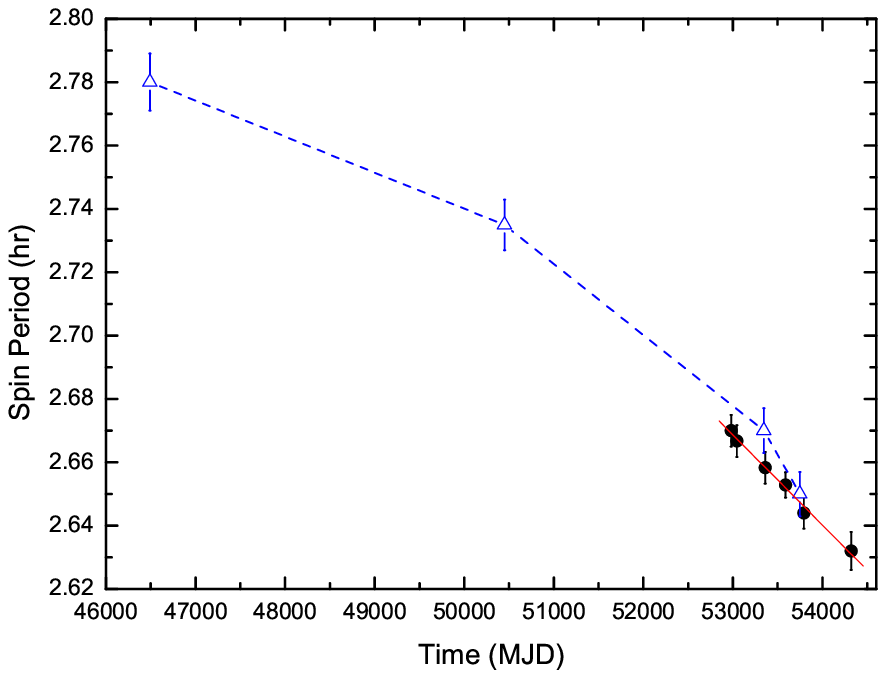}
\caption{Spin evolution history of 4U 2206+54 (left) and 2S 0114+65 (right) from different
observations in the last 20 years. }
\end{figure}

The long-term monitoring observations also discovered the spin evolution of these superslow pulsation X-ray pulsars. In Fig.\ 3, we have presented the spin evolution of 4U 2206+54 and 2S 0114+65 respectively. The pulsar in 4U 2206+54 undergone a long-term spin-down trend in the last twenty years with an average spin-down rate of $4.9\times 10^{-7}$\,s s$^{-1}$ by different measurements (Wang 2010, 2012; Finger et al.\ 2010; Reig et al.\ 2012). But the spin period of the neutron star in 2S 0114+65 varies from 2.73\,hr around 1986 to 2.63\,hr around 2008 (Wang 2011) with the present spin-up rate of $1.09\times 10^{-6}$\,s s$^{-1}$. Additionally, the spin-up rate of the neutron star in 2S
0114+65 seems to be accelerating (see Fig.\ 3). A slow rotation neutron star of $P_{\rm spin}\sim 1062$\,s was also discovered in a Be X-ray transient SXP 1062 (Henault-Brunet et al.\ 2012). During a giant outburst, a very fast spin-down rate of $\sim 3\times 10^{-6}$\,s s$^{-1}$ is discovered in this X-ray pulsar (Haberl et al.\ 2012).

\section{Accreting Magnetars - A new class of neutron star systems?}

Discovery of these superslow pulsation X-ray pulsars provides the challenge to the present evolution model of X-ray binaries. Then what is physical origin for long spin period neutron stars?

Li \& van den Heuvel (1999) have suggested that neutron star spins down to the spin period range longer than 1000 s if the neutron star
was born as a magnetar with an initial magnetic field $\geq 10^{14}$ G. This ultra-strong magnetic field could decay to the normal value ranges of $10^{12}-10^{13}$ G within a few million years, so that superslow pulsation X-ray pulsars may be defined as magnetar descendants.

The alternative suggestion proposed by Ikhsanov (2007) shows that an additional evolution phase {\em subsonic propeller} state between the transition from known {\em supersonic propeller} state to {\em accretor} state could allow the spin period increases up to several thousand seconds without the assumption of magnetars, which is the so-called break period given by: \beq  P_{\rm br} \simeq 2000 {M_{\rm NS}\over 1.4M_\odot}^{-4/21}[{B_{\rm surf}\over 0.3B_{\rm cr}}]^{16/21}[{\dot M \over 10^{15} {\rm g s^{-1}}}]^{-5/7} {\rm s}, \enq where $B_{\rm surf}$ is the surface magnetic field of the neutron star. However, if the above formula is applied to the case of 4U 2206+54/2S 0114+65, one find the surface magnetic field higher than $10^{14}$ G.

The fast spin-down rate is discovered in two superslow pulsation X-ray pulsars 4U 2206+54 and SXP 1062. According to the standard evolutionary scenario, the maximum spin-down rate in the accretor stage is $\dot P\sim 2\pi B^2R^6_{\rm NS}/(GMI)$, which implies $B> 10^{14}$ G for 4U 2206+54 and SXP 1062.

Recently, a new theory of quasi-spherical accretion for X-ray pulsars is developed (Shakura et al.\ 2012), the magnetic field in wind-fed neutron star systems is given by \beq B_{12}\sim 8.1\dot M^{1/3}_{16}V^{-11/3}_{300}({P_{1000}\over P_{orb300}})^{11/12} G. \enq This also gives the ultrastrong magnetic field of $>10^{14}$ G.

Thus, these superslow pulsation pulsars could be accreting magnetars! It is still quite interesting that the discovery of the cyclotron absorption line feature around 30 keV would suggest a magnetic field of 3$\times 10^{12}$ G for the electron cyclotron absorption case, but a magnetic field of $\sim 5\times 10^{15}$ G for the proton cyclotron line assumption. Thus, difficulty and uncertainties in explaining the long spin period still exist. It is possible that superslow pulsation X-ray pulsars may not follow the present standard evolution models in close
binaries.

The superslow pulsation X-ray pulsars undergo the fast spin evolution, not reaching the equilibrium. We suggested the possible evolution track among the superslow pulsation X-ray pulsars and supergiant binaries. Superslow pulsation X-ray pulsars should be younger binary systems, and after rapid spin-down and spin-up phases, they will become supergiant X-ray binaries in the equilibrium spin-period range.


\end{document}